\documentstyle[epsfig]{aipproc}

\begin{document}

\title{Prospects for radio detection of extremely high energy cosmic
rays and neutrinos in the Moon}

\author{Jaime Alvarez-Mu\~niz$^*$ and Enrique Zas$^{\dagger}$}
\address{$^*$Bartol Research Institute, University of Delaware, Newark,
DE 19711
\\
$^{\dagger}$Departamento de F\'\i sica de Part\'\i culas, 
Universidade de Santiago de Compostela,\\ 
E-15706, Santiago
de Compostela, Spain}

\maketitle

\begin{abstract}
We explore the feasibility of using the Moon as a detector
of extremely high energy ($>10^{19}$ eV) cosmic rays and neutrinos. 
The idea is
to use the existing radiotelescopes on Earth to look for 
short pulses of Cherenkov radiation in the GHz range emitted
by showers induced just below the surface of the Moon when 
cosmic rays or neutrinos strike it. We estimate the energy
threshold of the technique and the effective aperture and 
volume of the Moon for this detection. 
We apply our calculation to obtain 
the expected event rates from the observed cosmic ray flux and several
representative theoretical neutrino fluxes.   

\end{abstract}

\section*{Introduction}

The observation of atmospheric showers with total energy 
above the so-called GZK cutoff at energies above $\sim 5\times 10^{19}$
eV, is one of the most puzzling mysteries of the emerging 
field of astroparticle 
physics \cite{watson}. The nature of the primaries is still unknown as well as 
their origin and the mechanisms involved in their acceleration 
to such energies. The puzzle is even more intriguing because
if the primaries are protons, photons or nuclei they should be produced
within distances of a few tens of megaparsecs from us, otherwise
their interactions with the 2.7 K photons constituting the Cosmic Microwave
Background (CMB) as well as with the infrared and radiobackgrounds, 
should greatly reduce their energy. Since 
no effective sources capable of accelerating them have been identified
within that distance, their very presence is contrary to the expectations.
Neutrinos may play a fundamental role in solving this puzzle. They 
may be produced in interactions of the primary cosmic rays with the 
matter and radiation in the source
or in the matter and radiation they find along their paths to Earth.
The so-called GZK neutrinos produced in interactions of cosmic rays
with the CMB are almost guaranteed, since both the projectile and the 
target are known to exist \cite{steckerCMB,engel}.
Sources in which 
ultra high energy neutrinos may be produced include AGN's and 
GRB's \cite{halzennu} where accelerated particles interact with matter
or radiation. 
%
%Their expected fluxes are normalized assuming they are the sources
%of primary cosmic rays which leads to certain bounds 
%in their emission \cite{wblimit}. 
%One should however keep in mind that unbound neutrino sources may 
%exist in the Universe which are not expected to be cosmic ray 
%emitters due to their high optical depth which absorbs  
%ultrahigh energy protons, nuclei and photons but probably not neutrinos.  
%
Alternative sources are the annihilation of topological defects 
which can be produced
in phase transitions in the early 
universe where extremely high densities of energy are 
trapped \cite{sigl}. 
%
%Their decays or annihilations can produce GUT scale 
%particles of masses ${\rm M_X}\sim 10^{14}-10^{15}$ GeV which 
%at the same time decay producing the observed cosmic rays.
   
Some present and future detectors of ultra high energy cosmic rays 
\cite{watson,auger,owl,euso} and neutrinos \cite{halzennu,physrep,learnednu} 
will soon remedy the lack of statistics \cite{zaspuebla}. 
The small observed (predicted) fluxes of ultra high energy cosmic rays 
(neutrinos) call for huge effective volumes (measured in ${\rm km^3}$). 
Several detection methods and 
techniques are being employed and/or studied to achieve 
the required volumes, some of them have  
%
%Optical and radio 
%Cherenkov detection, the radar technique, acoustic signals, 
%atmospheric fluorescence, observation of horizontal showers 
%\cite{cronin} and others, have 
%
been discussed in this workshop \cite{RADHEP2000}.   
Among them the radio technique is one
of the most promising alternatives for neutrino and possibly cosmic
ray detection at ultra high energies ($10^{15}$ eV and above).
The technique aims at detecting {\it coherent} 
Cherenkov radiation in the MHz-GHz range from the excess of electrons
in showers initiated by photons or electrons. 
Several detectors are being planned \cite{RADHEP2000} or  
they are in the early stages of construction \cite{RICE}. Due to its 
excellent radio frequency wave propagation properties, Antarctic 
ice is being considered as a medium where 
antennas are being placed to monitor the potential radio signals.
Other media such as salt \cite{chiba} and sand are also being studied. 
As an interesting alternative,
in 1989 Zheleznykh et al. \cite{zheleznykh} proposed to detect showers
initiated by cosmic ray and neutrinos by measuring coherent Cherenkov
radiation emitted just below the surface of the Moon when high
energy cosmic rays or neutrinos strike it. Two groups have 
looked for those signals using existing radiotelescopes on Earth
with no positive detection during the 
time they have pointed the instruments to the Moon 
\cite{hankins,gorham}.     

Here we investigate the feasibility of detecting extremely high energy
particles interacting in the Moon. We discuss the characteristics of the radio
signals and we stress how they influence the aperture of the Moon 
as a high energy particle detector. We calculate the energy threshold
for the technique as a function of the detector parameters and the 
signal's main features. Using several representative neutrino fluxes,
we estimate the expected event rates and we briefly discuss where,
on the surface of the Moon, should the radiotelescopes be pointed at 
to be able to detect them. We also discuss  
cosmic ray detection. 
This work updates previous estimates in which some approximations to 
extremely high energy shower development in the lunar rock 
as well as to the associated radiosignals were
used \cite{jaimemoon}. We also take into account 
the transmissivity of the signals through the Moon's surface and some other 
geometrical issues which were not fully considered in our 
previous estimates.   

\section*{Radiopulses in the Moon}

Neutrino interactions produce different types of showers depending
on the neutrino flavor and on whether the interaction
is mediated by a ${\rm W}^\pm$ or a Z boson. 
In deep inelastic scattering (DIS) charged current (CC) interactions of 
electron neutrinos ($\nu_e$), the electron produced in
the lepton vertex initiates an electromagnetic shower
of energy $(1-y){\rm E_\nu}$, where ${\rm E_\nu}$ is the neutrino
energy in the laboratory frame and $y$ is the fraction of energy 
transferred to the struck nucleon
in the hadronic vertex. The debris of the nucleon initiate a hadronic 
shower which is superimposed to the electromagnetic one producing a 
``mixed shower''. Hadronic showers are also initiated in both charged 
and neutral current (NC) DIS neutrino interactions by the hadrons 
created in the fragmentation of the nuclear debris. 
%
%In this work we are only 
%going to consider these two types of showers. Purely electromagnetic
%showers may also be initiated in the decays of ${\rm W}^+$ around the 
%energy where the reaction $\bar\nu_e+e^-\rightarrow {\rm W}^+$ has 
%a resonance ($\sim 6.4$ PeV), however this 
%energy is well below the estimated threshold for observation
%from the Moon (see below) and hence we will not consider it 
%in our estimates. 
%
At shower energies above $E_{\rm LPM}\sim 4 \times 10^{14}$ eV (400 TeV) 
in the lunar regolith, electromagnetic showers are affected by the 
LPM effect \cite{LPM}. When the energy of a photon or electron in the 
shower is above $E_{\rm LPM}$ the interaction distance becomes 
comparable to the interatomic spacing and collective atomic 
and molecular effects affect the static electric
field responsible for the interaction. The result is a reduction in both
total cross sections with lab energy ($E$) which drop like 
$E^{-0.5}$ above $\sim E_{LPM}$.
As a result PeV - EeV photon or electron induced showers are 
considerably larger than at TeV energies \cite{Alz97}.
In hadronic showers this effect is mitigated 
because the main source of photons  
is the decay of $\pi^0$'s which at energies above 
$E_{{\pi}^0}=7 \times 10^{15}$ eV (7 PeV) is suppressed due to 
the dominance of $\pi^0$ interaction \cite{Alz98}.  

A helpful way of  
getting insight into the pattern of the electric field emitted by 
the excess of electrons in the shower, is to
think of a shower as a slit illuminated by radio-waves \cite{Alz99}. 
The diffraction pattern can be obtained as the Fourier transform 
of the slit, however due to the nature of the Cherenkov radiation 
the emission has a central peak with its maximum 
at the Cherenkov angle instead of 
in the direction perpendicular to the slit. The radiation
is coherent when the wavelength is larger 
than the physical dimensions of the shower.   
In this situation the radiated electric field 
becomes proportional to the excess charge 
and the power in radiowaves scales
with the square of the shower energy. This effect
was predicted in the 1960's by Askary'an \cite{aska} and has
been observed recently at SLAC in experiments where an intense beam  
of GeV photons generates a shower inside a sand target \cite{saltzberg}. 
Given this interpretation
it is easy to understand that the LPM effect is going to 
reduce the angular width of the Cherenkov peak ($\Delta\theta$) 
in electromagnetic showers as shower energy increases. Following the 
same reasoning, the diffraction pattern from hadronic showers is
expected to be less affected by the LPM, exhibiting a mild decrease
of $\Delta\theta$ with energy \cite{Alz98}.   

\begin{figure}[hbt!] % fig 1
\centerline{\epsfig{file=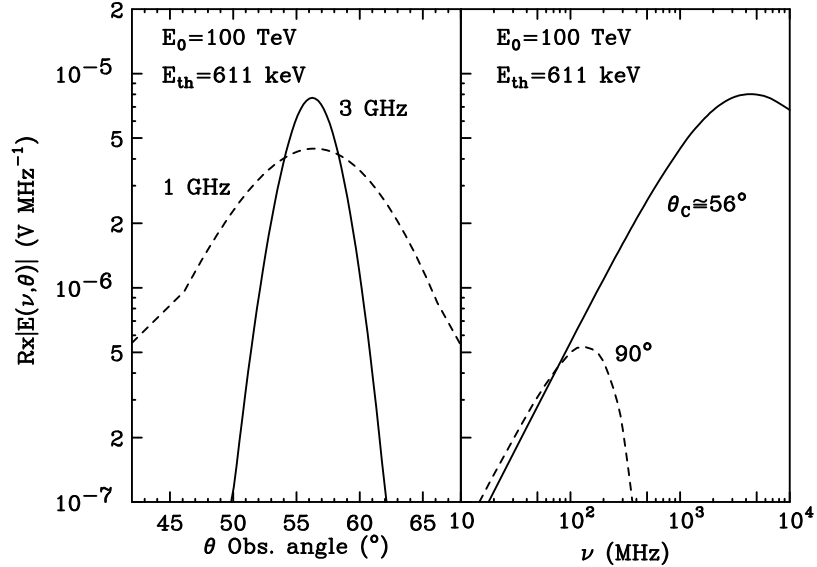,height=3.0in}}
\vspace{10pt}
\caption{Left panel:  Results of simulations of the angular 
behavior of the electric field 
emitted by a 100 TeV electromagnetic shower in the lunar regolith. 
$\theta$ is the observation angle with
respect to the shower axis. Two frequencies are shown. Right panel: Frequency 
spectrum of the electric field at two observation angles.}
\label{fig1}
\end{figure}

The spectrum of the electric field also increases linearly 
with frequency up to 
a maximum frequency which is essentially determined by the lateral
structure of the shower at the Cherenkov angle, the wider the shower
the smaller the corresponding maximum frequency. At angles away
from the Cherenkov angle, the maximum frequency depends
only on the longitudinal dimension of the shower. 
We have performed simulations of electromagnetic showers 
in the lunar regolith assumed homogeneous (density 
$\rho\simeq 1.6~{\rm g~cm^{-3}}$). We have adapted 
the ZHS code \cite{zas92} originally conceived for simulations 
in ice, changing the relevant parameters such
as density, radiation length, atomic number and refraction index. 
The frequency spectrum
as well as the angular distribution of the pulses emitted by a 
100 TeV electromagnetic 
shower in the Moon are shown in Fig.\ref{fig1}. The results
of the simulations are in very good agreement with a simple scaling
of the simulations from ice to the Moon \cite{jaimemoon}. 
%
%we have chosen not to perform the time consuming simulations
%of ultra high energy showers in the Moon 
%
With this in mind we have obtained the 
electric field from electromagnetic showers at the highest energies, 
by scaling the spectral features in ice.
%
%of the spectrum in ice 
%but normalizing the electric field according to the result obtained 
%from the simulations in the Moon. 
The same scaling has been
applied to estimate the radiopulses from hadronic showers. 
The absolute
value of the electric field has been normalized according to the 
amount of electromagnetic energy present in the hadronic shower, and
the width of the Cherenkov peak, which is inversely  
proportional to the longitudinal
dimension of the shower, is scaled with the radiation length and 
the density of the medium \cite{jaimemoon}. Simulation
work is in progress in this direction and will be published elsewhere.
For more details see references \cite{Alz97,Alz98}. 

\section*{Aperture of the Moon for neutrino and cosmic ray detection:
Energy threshold}

\begin{figure}[hbt!] % fig 2
\centerline{\epsfig{file=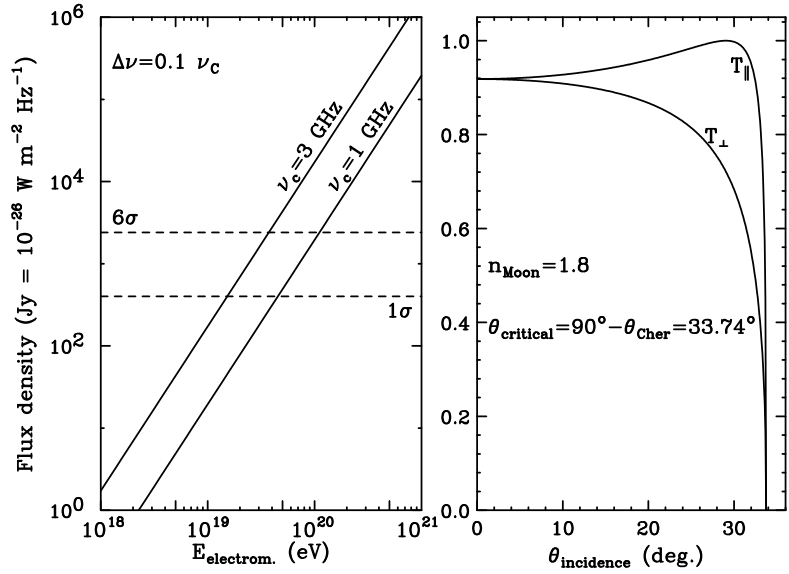,height=3.0in}}
\vspace{10pt}
\caption{Left panel: Flux density at Earth emitted by an electromagnetic
shower of energy $E_{\rm electrom.}$ for two central frequencies and 
a bandwidth $\Delta\nu=0.1 \nu_C$. Also shown is the flux density of the 
thermal noise in NASA/JPL Goldstone DSS 14 radiotelescope. Right panel:
Transmissivity (${\rm Power_{transmitted}/Power_{incident}}$)
of radio waves at the Moon-vacuum interface as a function of the 
incidence angle with respect to the perpendicular to the local surface
of the Moon.}
\label{fig2}
\end{figure}

One important parameter for assessing the possibilities of using the Moon
as a high energy particle detector is the energy threshold 
for shower observation by radiotelescopes on Earth. This can be 
estimated comparing the radio signal emitted by an electromagnetic shower
at the Cherenkov peak and the noise expected by other processes in 
the radiotelescope. Here we only take into account thermal noise for 
which we use two representative values, 
namely $1\sigma$ and $6\sigma$
of the flux density given by the standard expression for a radio antenna:

\begin{equation}
F_{\rm Noise}={2 k_{\rm B} T_{\rm sys} \over 
\sqrt{\Delta {\rm t} \Delta \nu} A_{\rm eff}}.
\end{equation} 
Here $k_{\rm B}$ is Boltzmann's constant, $T_{\rm sys}$ 
is the noise temperature of the radio detection system, $\Delta t$ and 
$\Delta \nu$ are respectively the duration of the pulse  
and the bandwidth of the detection system around a central 
frequency $\nu_C$. $A_{\rm eff}$ is the 
effective area of the antenna. We have estimated the energy threshold
for the NASA/JPL Goldstone Deep Space Station 14 (70 m diameter dish) 
radio telescope \cite{goldstone} with which the
measurements in \cite{gorham} were performed. 
The $1\sigma$ thermal noise level for this system is $\sim 400$ Jy 
\footnote{$1~{\rm Jy} = 10^{-26}~{\rm W~m^{-2}~Hz^{-1}}$} and 
the $6\sigma$ is 2,400 Jy. 
The flux density on Earth emitted by an electromagnetic shower 
at the Cherenkov peak is obtained integrating the frequency 
spectrum of the electric field 
${\rm R} {\bf E (\nu, {\rm R}, \theta_{Cher})}$, where ${\rm R}$ is 
the distance to the shower, around $\nu_{\rm C}$:

\begin{equation}
F_{\rm Signal}=\int_{\nu_C + {\Delta\nu\over 2}}^{\nu_C -
{\Delta\nu\over 2}}~\vert{\rm R} {\bf E (\nu, {\rm R}, \theta_{Cher})} 
\vert^2~d\nu~/~(4\pi {\rm R_{Earth\rightarrow Moon}^2}).
\end{equation} 
$ {\rm R_{Earth\rightarrow Moon}}$ is the distance from
Earth to the Moon. 
The result of this calculation for electromagnetic showers is 
shown in Fig.\ref{fig2}. 
Reading from the plot one can estimate the energy threshold of 
a electromagnetic shower to be between 
$E_{\rm th}\sim 10^{19}-10^{20}$ eV depending
on the noise level, the central frequency and assuming a bandwidth
of $\sim 0.1~\nu_C$. 
To translate the electromagnetic shower energy threshold 
into a neutrino or cosmic ray energy threshold, one has to 
keep in mind that at extremely high energy the electromagnetic
energy content of a hadronic shower in the lunar regolith 
is $\sim 90\%$ of the total shower energy, slowly  
increasing with it \cite{Alz98} and the 
typical value of $y$ is $\sim 0.2$ \cite{reno}. This implies that the 
energy threshold for detecting a NC neutrino DIS
interaction is roughly $E_\nu \sim 5 E_{\rm th}$ whereas it is 
$\sim E_{\rm th}$ for detection of CR or for a CC $\nu_e$ interaction. 

\subsection*{Aperture}

We have studied in detail the geometry of the problem in order
to calculate the aperture (effective area x solid angle) of the Moon
for neutrino as well as cosmic ray detection.  

\begin{figure}[hbt!] % fig 3
\centerline{\epsfig{file=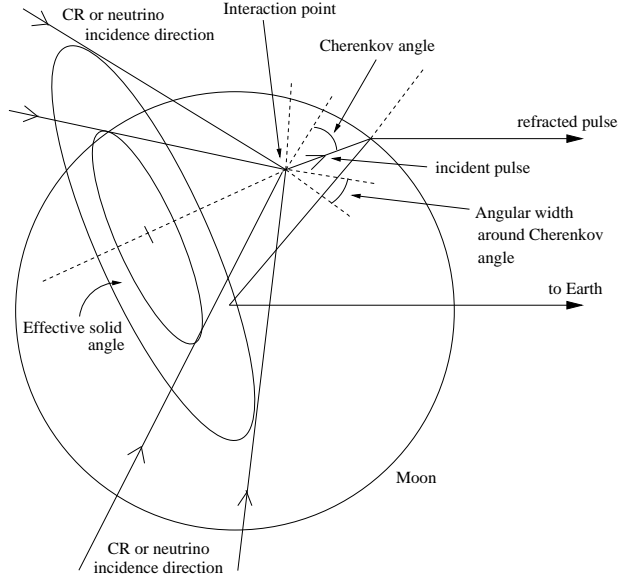,height=3.0in}}
\vspace{10pt}
\caption{Main elements entering in the geometry of 
radio detection in the Moon. All the cosmic ray and neutrino directions entering
in the depicted effective solid angle are ``geometrically allowed'',
although many of them are disregarded due to absorption inside
the Moon or threshold effects (see text).}
\label{fig3}
\end{figure}

\begin{figure}[hbt!] % fig 4
\centerline{\epsfig{file=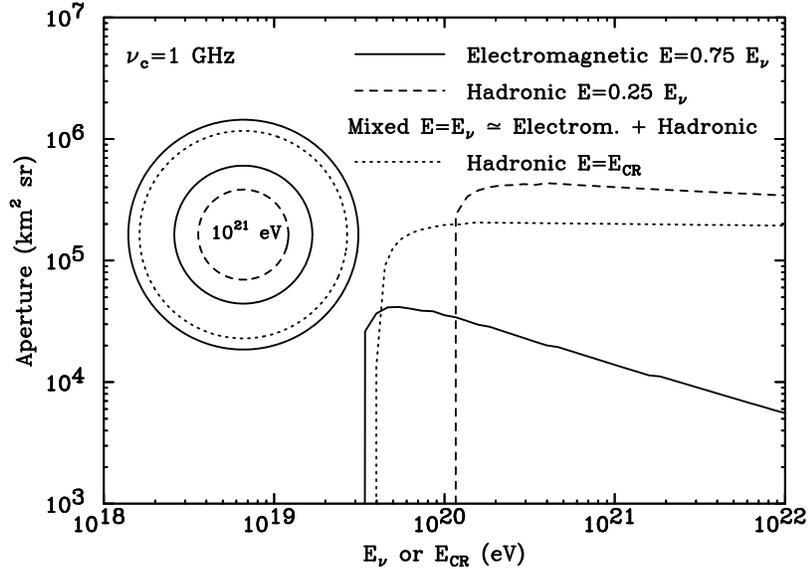,height=3.0in}}
\vspace{10pt}
\caption{Aperture of the Moon for cosmic ray and neutrino detection. 
For neutrino initiated mixed showers the aperture is the sum of 
the solid and dashed curves (not drawn). The inset represents the 
surface of the Moon. The lines correspond to the same lines in the
legend. The region between each line and the outermost solid line
is the surface where cosmic ray or neutrino detection is expected
at $E=10^{21}$ eV.}
\label{fig4}
\end{figure}

For a cosmic ray or a neutrino to be detected it has to interact 
producing a shower within a distance to the surface of the Moon equal to the 
typical absorption length of radiowaves inside the lunar regolith: 
$\lambda_{\rm abs}^{\rm radio}\sim 15~{\rm m} (1~{\rm GHz}/\nu)$.  
The emitted radiowave then reaches the surface of the Moon without
significant attenuation and has to 
point to the Earth after refraction in the Moon-vacuum interface.  
We restrict ourselves to radio emission from 
the directions between the Cherenkov angle and $\theta_C\pm\Delta\theta$, 
where most of the power is concentrated. These two considerations allowed 
us to obtain the potentially detectable incident neutrino and cosmic ray 
directions. Then we disregarded those directions for which either 
the cosmic rays or neutrinos are absorbed inside the Moon before 
reaching its surface\footnote{The interaction length ${\rm L_{int}}^\nu$ 
of a neutrino 
is smaller than the diameter of the Moon when $E_\nu > 10$ PeV}. 
It requires a little bit of thinking and the aid of Fig.\ref{fig3} 
to realize that this limits the potential interaction points on the 
surface of the Moon to a fairly narrow rim for cosmic
rays and to a wider rim for neutrinos. It also restricts the observed 
cosmic rays to those hitting the Moon with their tracks almost parallel 
to its surface. 
We finally eliminated those
directions (and energies) 
for which the signal, although being ``geometrically
able'' to reach Earth, is below the noise flux density. 
In this respect the transmissivity properties of the Moon-vaccumm interface
play a fundamental role.  
For a shower propagating inside the Moon parallel to 
the local surface, the radiation emitted at the Cherenkov angle suffers
total internal reflection since the Cherenkov angle is the complementary 
of the total internal reflection angle. However detection is still 
possible because, as we mentioned above, the radiation is emitted in 
a cone of width $\Delta\theta$ and the transmissivity of the 
interface increases to almost 1 just a few degrees off the total internal 
reflection angle as can be seen in Fig.\ref{fig2}. Our results 
on the aperture are shown in Fig.\ref{fig4}. 
It is interesting to 
notice that for hadronic showers once the energy 
threshold is reached, the aperture varies slowly with shower energy, the 
reason for this being that the effective solid angle is not  
restricted by an energy decrease of $\Delta\theta$ caused by  
the LPM effect. This is not the case for electromagnetic 
showers where the decrease of the aperture with energy 
is induced by a corresponding
reduction in $\Delta\theta$ due to the LPM.  
The inner panel in Fig.\ref{fig4}
represents the visible face of the Moon and gives an idea of the size 
of the rim where cosmic ray and neutrino detection might be expected
at $10^{21}$ eV.  
When combining the aperture with $\lambda_{\rm abs}^{\rm radio}$ acceptances 
in excess of $1000~{\rm km^3~sr}$ at $E_\nu\sim 10^{20}$ eV 
might be achieved. 

\section*{Event rate estimates}

Cosmic ray event rates are calculated in a straightforward manner
by convoluting the experimentally observed flux and the aperture of 
the Moon. For neutrinos the convolution involves the theoretically 
expected neutrino flux, the neutrino cross section and the effective volume of
the detector:

\begin{equation}
N_{\rm CR}=\int_{E_{\rm CR}}~{d\Phi_{\rm CR}\over dE_{\rm CR}}~
S_\Omega^{\rm CR}~dE_{\rm CR}~~{\rm (yr^{-1})};~~{\rm for~cosmic~rays,}
\end{equation}
\begin{equation}
N_\nu=\int_{E_\nu}~{d\Phi_\nu\over dE_\nu}~
S_\Omega^\nu~{\lambda_{\rm abs}^{\rm radio}\over {\rm L_{int}}^\nu}
dE_\nu~~{\rm (yr^{-1})};~~{\rm for~neutrinos,}
\end{equation}
where $S_\Omega^{CR,~\nu}$ is the aperture of the detector.  

\begin{figure}[hbt!] % fig 5
\centerline{\epsfig{file=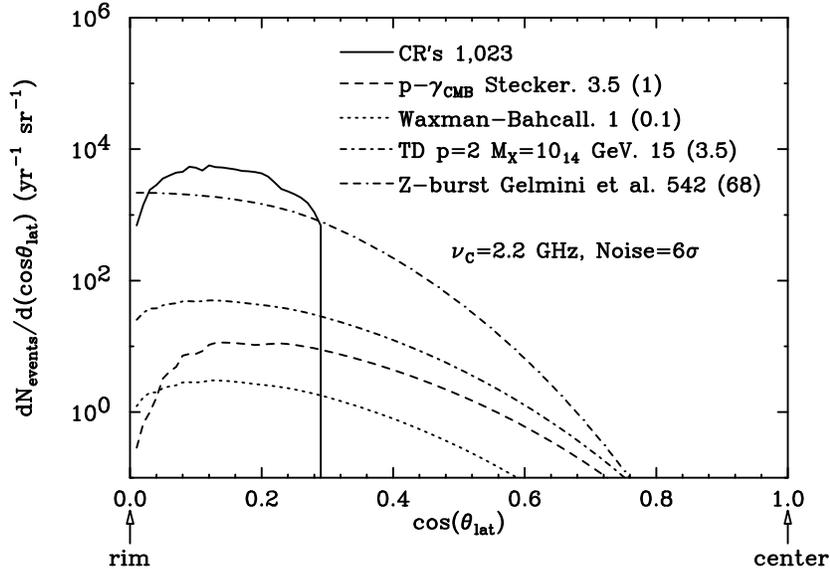,height=3.0in}}
\vspace{10pt}
\caption{Cosmic ray and $\nu_e$ CC + $\nu_e$ NC + $\nu_\mu$ CC+NC events
in the Moon expected in the NASA/JPL Goldstone DSS 14 radiotelescope. 
The events are shown
as a function of the cosine of the latitude of the points on the 
surface of the Moon,
namely $\cos(\theta_{\rm lat})=1$ corresponds to the center
of the Moon and $\cos(\theta_{\rm lat})=0$ to the outermost
rim. One should understand this plot as having azimuthal symmetry around 
$\cos(\theta_{\rm lat})=1$. The numbers accompanying the legends 
are the total events
per year. The numbers in parenthesis represent the neutrino events
per year from the latitudes where cosmic ray events are 
not expected. For cosmic rays the number of events below $3\times
10^{20}$ eV is shown.}
\label{fig5}
\end{figure}

In Fig.\ref{fig5} we show the event rates 
for the conservative case where the noise level is equal 
to $6\sigma$ of the thermal noise. 
We have used some neutrino fluxes representative of 
the different current models and theoretical ideas.  
For the cosmic ray event rate, we have used a simple parameterization
of the experimentally observed flux which we
have extended, perhaps conservatively, to the highest energy 
observed event ($3\times 10^{20}$ eV) 
using a differential slope $\gamma\sim-2.7$.
It is clear from the figure that cosmic ray events are going to dominate
the event rate near the rim of the Moon outnumbering the neutrino
events at least for the representative neutrino fluxes we have chosen. 
It is then advisable to
point the radiotelescopes towards the center of the Moon 
to be able to observe neutrino events. 
Due to their high angular resolution, the radiotelescopes such as 
Goldstone DSS 14 or Parkes in Australia, are able to observe 
roughly $10\%$ of the surface of the Moon at once, hence
one should multiply the numbers in Fig.\ref{fig5} by a factor of 0.1, to get 
an estimate of the sensitivity of the existing instruments and the 
observation time required to collect a certain number of events.  
Our estimates indicate that Goldstone DSS 14 should roughly expect 
1 cosmic ray event every 80 hours of observation. 

\section*{Improvements and conclusions}

There are still a few issues  
that should be explored.  
The roughness of the surface of the Moon should be included in 
the calculation. Its wrinkles may facilitate the detection
of signals emitted at the Cherenkov angle since, in this case, 
the particle tracks skimming the regolith surface are in general
not parallel to the local surface which creates 
a problem of total internal reflection.   
Since showers should be produced near the surface
of the Moon, one should worry about near-field effects in the 
Cherenkov emission \cite{buniy}. 
Their main consequences are a decrease of the normalization of 
the electric field at the Cherenkov peak accompanied by  
an increase of the width of the Cherenkov cone $\Delta\theta$
\cite{Alz00}. These two effects act in opposite directions
with respect to the behavior of the event rate. A decrease in 
$\Delta\theta$ will increase the effective solid angle while 
the decrease in the normalization increases the energy threshold.  
There are also some issues dealing with showers
produced at a distance to the surface of the Moon smaller than
a radiation length so that Cherenkov radiation doesn't have enough 
distance to form. Variations of the index of refraction near 
the surface of the Moon have been measured and should also 
be included in the calculation of the geometry. 
Work is in progress to estimate the importance of all these 
potential issues and the results will be published elsewhere. 

Although some room for improvement does exist, we have identified
and correctly taken into account many of the elements entering 
in this difficult problem, updating previous estimates \cite{jaimemoon}. 
The actual potential of the Moon as a 
detector of high energy cosmic rays and neutrinos depends on 
a comprehensive study of all the elements indicated in this 
paper and possibly others. In any case,
our results are encouraging and should trigger more theoretical
work on this exciting possibility. 
 
\section*{Acknowledgements}

We thank Ricardo V\'azquez for many discussions
on this work. 
We thank P. Gorham and D. Saltzberg 
for an excellent and stimulating workshop.
We acknowledge the suggestions 
made by some of the participants: D. Besson, G. Gelmini, 
P. Gorham, F. Halzen, K. Liewer, D.W. McKay, C. Naudet, 
J.P. Ralston, D. Saltzberg and I. Zheleznykh among
others. The research activities of J.A-M at Bartol Research Institute 
are supported by the NASA grant NAG5--7009.  
This work was also supported in
part by the European Science Foundation 
(Neutrino Astrophysics Network N. 86), by the CICYT 
(AEN99-0589-C02-02) and by Xunta de Galicia (PGIDT00PXI20615PR).

\end{document}